\documentclass[prd,preprintnumbers,floatfix,
nofootinbib,superscriptaddress, two column, linenumbers]{revtex4}

\usepackage{mathtools}
\usepackage{graphicx}
\usepackage{hyperref}

\newcommand{\diff}{\mathrm{d}}
\newcommand{\ket}[1]{\ensuremath{\left|#1\right\rangle}}
\newcommand{\bra}[1]{\ensuremath{\left\langle #1\right|}}

\newcommand{\TS}{\mathrm{TS}}
\newcommand{\LLH}{\mathcal{L}}

\newcommand{\I}{\ensuremath{I}}
\newcommand{\II}{\ensuremath{{I\!I}}}
\newcommand{\III}{\ensuremath{{I\!I\!I}}}
\newcommand{\IV}{\ensuremath{{I\!V}}}
\graphicspath{{figs/}}

\begin{document}
\title{The determination of the spin and parity of a vector-vector system}

\author{Liupan An}
\affiliation{School of Physics, Peking University, Beijing, China}
\author{Ronan McNulty}
\affiliation{School of Physics, University College Dublin, Dublin 4, Ireland}
\author{Mikhail Mikhasenko}
\email[e-mail: ]{mikhail.mikhasenko@cern.ch}
\affiliation{Ruhr University Bochum, Universitätsstraße 150, Bochum, Germany}

\date{\today}

\begin{abstract}
  We present a construction of the reaction amplitude
  for the inclusive production of a resonance decaying to a pair of identical vector particles such as $J/\psi J/\psi$, $\rho\rho$, $\phi\phi$.
  The method provides the possibility of determining the spin and parity of a resonance in a model-independent way.
  The methodology is demonstrated using the Standard Model decay of the Higgs boson to four leptons and through angular correlations in the production and decays of $J/\psi$ pairs.
\end{abstract}

\nopagebreak
\maketitle

\section{Introduction}

The formation of hadronic matter is one of the few poorly understood parts of Quantum Chromodynamics (QCD).
The quark model (QM)~\cite{GellMann:1964nj,Godfrey:1985xj} works well in classifying conventional hadronic states into mesons and baryons built from the constituent quarks bound in the confined potential.
Hadrons beyond conventional mesons and baryons, such as glueballs, hybrid states, and multiquark states, are referred to as exotic hadrons~\cite{Klempt:2007cp,Meyer:2015eta}.
They are allowed by the QM, however they have not been seen experimentally until recently~\cite{Olsen:2017bmm, Brambilla:2019esw}.
Over the last decade, overwhelming evidence
has accumulated for exotic hadrons including
tetraquarks, pentaquarks and hadronic molecules in
the heavy quarkonium system~\cite{Godfrey:2008nc, Aaij:2015tga,Aaij:2019vzc,Guo:2017jvc}, and beyond~\cite{LHCb:2020bls,LHCb:2020pxc,LHCb:2020bwg,LHCb:2022sfr,LHCb:2022lzp,LHCb:2021vvq,LHCb:2021auc}.
Despite the large progress in the field,
the overall picture and the categorisation of these states remain unclear.

While the quark content of the exotic state is often straightforwardly determined from its decay channel,
the quantum numbers are rarely learned in a simple analysis of the mass spectrum.
The spin and parity of a resonance are usually found from angular correlations of the decaying particles.
One approach to relating such correlations to properties of particles is to derive the reaction matrix element from the interaction terms in the Lagrangian.
For non-perturbative interaction of hadrons, an alternative technique is required.
We explore the approach that relies on the fundamental constraints and symmetries appurtenant to the strong-interaction theory with no assumption on the underlying dynamics.
The reaction amplitude is built using the helicity formalism~\cite{Jacob:1959at}, where
the functional form of the matrix element follows from the rotational properties of the decaying particle.
The two key constraints that determine the decay properties are parity conservation and permutation symmetry.

In this paper, we provide a practical implementation of the helicity formalism to
determine the spin-parity for
a system of two identical vector bosons that decay to a pair of leptons or a pair of scalar particles.
Amongst its many applications, we note three in particular.
First, it facilitates studies of the $J/\psi J/\psi$ system, where resonance-like structures were reported by the LHCb experiment~\cite{LHCb:2020bwg} and confirmed by both the ATLAS~\cite{ATLAS:2023bft} and CMS~\cite{CMS:2023owd} experiments.
In none of the three analyses is the interpretation of the $J/\psi J/\psi$ spectrum unambiguously determined.
The spectrum shows a broad structure above the $J/\psi J/\psi$ threshold, followed by a resonance-like peak with a dip in between.
This can variously be described as multiple resonances or interference between resonant and non-resonant components, and could have contributions from partially reconstructed decays of heavier state.
An angular analysis can help to distinguish different scenarios~\cite{Dong:2020nwy,Wang:2020wrp,Karliner:2020dta,Gong:2020bmg} and knowledge of the quantum numbers of the states can elucidate the mechanism for the binding of four charm quarks~\cite{Liu:2019zoy,Weng:2020jao,Chen:2020xwe,An:2022qpt,Becchi:2020uvq}.

The second application is the
investigations of the central exclusive production (CEP) of vector-meson pairs.
The colour-free gluon-rich production mechanism of CEP makes the
$\rho\rho$~\cite{Armstrong:1989jk, Abatzis:1994ym, Osterberg:2014mta} and
$\phi\phi$~\cite{Barberis:2000em,Lebiedowicz:2019jru}
channels particularly suited to searches for glueballs.
The helicity approach has been used before to analyse the $\phi\phi$ system decaying to four pseudoscalar particles by several authors~\cite{Chang:1978jb,Trueman:1978kh}.
Our method generalizes these ideas and provides an expression for the decay rate
as a function of decay angles for a particle with arbitrary spin.
It paves the way for a complete partial-wave analysis of the high statistics CEP of four scalar mesons that should be possible with modern LHC data.

The third study-case is the decay of Higgs to two $Z$ bosons and their subsequent
decay into four leptons.
This final state has been extensively studied in the past as the golden decay mode of the Higgs boson and in searches for physics beyond the Standard Model~\cite{Keung:2008ve,Gao:2010qx,Bolognesi:2012mm,Modak:2014zca,Berge:2015jra}.
We use this decay to validate the methodology and as an example of application of statistical methods to distinguish between different spin-parity hypotheses.

We consider the decay $X\to V(l^+l^-)V(l^+l^-)$ in the rest of this paper.
Modifications needed for vector decays to scalars $X\to V(S^+S^-)V(S^+S^-)$ are given in Appendix~\ref{sec:KKKK}.
The paper is organized as follows.
The theoretical formalism is presented in Sec.~\ref{sec:formalism}, split into three parts that
deal with:
the construction of the general reaction amplitude using the helicity formalism;
the implementation of symmetry constraints; and
the discussion of angular moments that can be experimentally observed.
This is followed by two applications of these equations in Sec.~\ref{sec:application}:
the Higgs decay as an example of an isolated resonance
and the $J/\psi J/\psi$ system as a more complicated case.
Conclusions follow in Sec.~\ref{sec:conclude}.


\section{Formalism} \label{sec:formalism}
\subsection{Angular amplitude} \label{sec:reaction.amplitude}
We focus on the inclusive production process $p p'\to X + \dots$, where $X$ is a resonance decaying to two vector mesons.
Although the vector mesons are identical, it is convenient to distinguish them in the reaction amplitude calling them $V_1$ and $V_2$.
In that way, we can make sure that the amplitude is symmetric under the permutation of indices $1$ and $2$.
When the decay modes of the two vectors are identical, namely \mbox{$X\to V(l_1^+l_1^-)V(l_2^+l_2^-)$},
one needs to account for the symmetrized process \mbox{$X\to V(l_1^+l_2^-)V(l_1^+l_2^-)$}
and the interference between the two decay chains.
For narrow resonances, the interference is minor and is neglected in the following discussion.
The calculation of interference effects lies beyond the scope of this paper.

\begin{figure}
  \includegraphics[width=\linewidth]{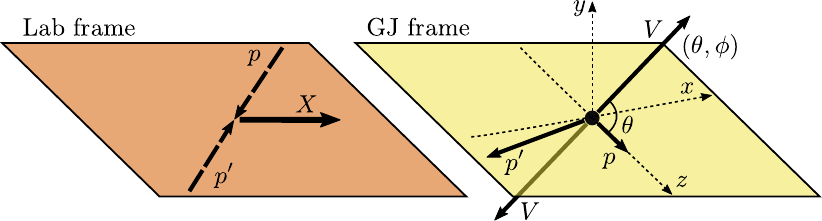}
  \caption{Schematic view of the production kinematics of the $X$ state in $pp$ collisions.
    The Gottfried-Jackson frame is used:
    the axes are defined in the rest frame of $X$ by the vectors of the beam particles:
    $\vec z = \vec p / |\vec p|$, $\vec y = \vec p' \times \vec p / |\vec p' \times \vec p|$, $\vec x = \vec y \times \vec z$.
    The spherical angles $(\theta,\phi)$ are the angles of one of the two decay vectors in the GJ frame.
    The black arrows shows the three-vectors of the particles.
    The three-momenta of the vector mesons are labeled by $V$.
  }
  \label{fig:production}
\end{figure}

The production frame is set up in the rest frame of $X$ as a plane that contains the three-vectors of the
production reaction, i.e.\  $\vec p$ and $\vec p\,'$. The normal to the plane gives the $y$ axis ($\vec p\,'\times \vec p$) as shown in Fig.~\ref{fig:production}.
The Gottfried-Jackson (GJ) frame is used to define the $x$ and $z$ axes in the production plane~\cite{Gottfried:1964nx},
where the $z$ axis is defined along the direction of $\vec p$.
The choice of $x$ and $z$ axes is not unique: two other common definitions of the production frame are the helicity (HX) frame, where the $z$ axis is
defined by the direction of motion of $X$ itself in the lab frame~\cite{Jacob:1959at}, and the Collins-Soper (CS) frame in which $z$ is defined by the bisector of the angle between $\vec p$ and $\vec p\,'$~\cite{Collins:1977iv}.
We note that a negligibly small polarization is measured in the prompt production of charmonium ("head on" collisions)~\cite{Aaij:2013nlm,Chatrchyan:2012woa,CDF:2011ag, Aaij:2013oxa,Sirunyan:2018bfd}.
In contrast, for peripheral processes, e.g. central exclusive production, a significant polarization is expected~\cite{Pasechnik:2010pq}.
Therefore we consider the general case of an arbitrary polarization of $X$.

\begin{figure*}
  \includegraphics[width=0.8\textwidth]{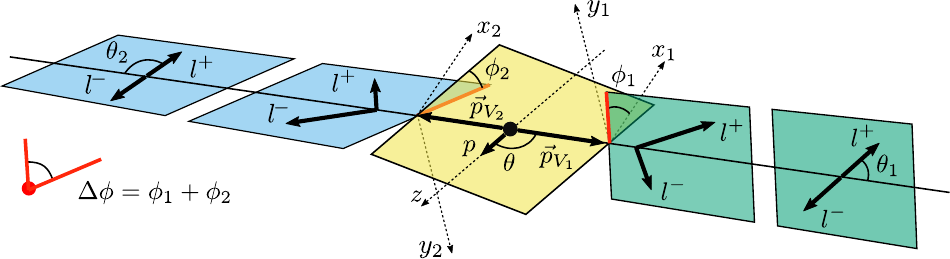}
  \caption{Schematic view of the $X\to V(l^+l^-)\,V(l^+l^-)$ decay kinematics.
    The central three planes show the orientation of the vector mesons in the $X$ rest frame.
    The rightmost plane shows the helicity angle $\theta_1$ for the $V_1$ decay in the rest frame of the two leptons while the leftmost plane shows the helicity for the $V_2$ decay in the rest frame of the other pair of leptons.
    The $x_1, y_1$ and $x_2, y_2$ axes indicate the directions with respect to which the azimuthal angles of the positive leptons are measured in the decays of the first and second vector, respectively. Both $x_1$ and $x_2$ belong to the central yellow plane, both $y_1$ and $y_2$ are orthogonal to it.
  }
  \label{fig:decay}
\end{figure*}

The full kinematics of the decay is described by six angles: a pair of spherical angles $\Omega = (\theta,\phi)$ of the momentum of $V_1$ in the GJ frame, and two pairs of spherical angles $\Omega_i = (\theta_i,\phi_i)$, $i=1,2$ for the decays of the vector mesons $V_i$ in their own HX frames,
as illustrated in Fig.~\ref{fig:decay}.
The coordinate system for the second vector meson is obtained by the rotation by $\pi$ about the $y$ axis, followed by a rotation by $\pi$ about the $z$ axis.
The angles $\phi_i$ can also be defined in the $X$ rest frame as shown in Fig.~\ref{fig:decay} since they are not affected by the boosts along the vector-meson directions of momentum.
The spin of the decay particle $X$ defines the rotational properties of the system of decay products~\cite{Mikhasenko:2019rjf}.
Every configuration of the three-momenta of the final-state particles in the $X$ rest frame
can be considered as a solid body for which the orientation is described by three angles:
the pair of spherical angles $(\theta,\phi)$ that describe the direction of $\vec p_{V_1}$,
and $\phi_1$, the azimuthal direction of $l^+$ (see Fig.~\ref{fig:production} and Fig.~\ref{fig:decay}).
The normalized differential cross section, denoted by the intensity $I$, reads:
\begin{align} \label{eq:I.6}
   & I(\Omega,\Omega_1,\Omega_2) = \sum_{J}(2J+1) \sum_{M,M'}R_{M,M'}\,                               \\ \nonumber
   & \quad\times\sum_{\nu,\nu'}D_{M,\nu}^{J*}(\phi,\theta,\phi_1) D_{M',\nu'}^{J}(\phi,\theta,\phi_1)
  \\ \nonumber
   & \qquad\times
  \sum_{\xi_1,\xi_2}^{\{-1,1\}}A^{J;\nu}_{\xi_1,\xi_2}(\theta_1,\theta_2,\Delta\phi) A^{J;\nu'*}_{\xi_1,\xi_2}(\theta_1,\theta_2,\Delta\phi),
\end{align}
where the production and decay parts of the amplitude are explicitly separated.
The spin of $X$ is denoted by $J$ and $M$ is its spin projection onto the $z$ axis.
The definition of the Wigner D-function can be found in Ref.~\cite{Collins:1977iv}.
The production dynamics are encapsulated in the polarization matrix $R_{M,M'}$.
The decay amplitude is denoted by $A^{\nu}_{\xi_1,\xi_2}$,
where $\nu$ is the difference of the vector-meson's helicities, $-2\leq \nu \leq 2$,
and $\xi_i$ is the difference of the two leptons' helicities in the decay of $V_i$, $-1\leq \xi_i \leq 1$, $i=1,2$.
As $\xi_i=0$ is suppressed by $m_l/m_{V}$ for the electromagnetic transition, we omit it in the summation.
The remaining $V\to l^+l^-$ helicity couplings give an overall constant.
The decay amplitude is described by the remaining three angles, $\theta_1$, $\theta_2$, and $\Delta\phi = \phi_2+\phi_1$ (see Fig.~\ref{fig:decay}), and is given by
\begin{align} \label{eq:decay.A}
   & A^{J;\nu}_{\xi_1,\xi_2}(\theta_1,\theta_2,\Delta\phi) = \\ \nonumber
   & \quad\frac{3}{2}
  \sum_{\lambda_1,\lambda_2}
  \delta_{\nu,\lambda_1-\lambda_2}
  H_{\lambda_1\lambda_2}^{J}
  d_{\lambda_1,\xi_1}^{1}(\theta_1) d_{\lambda_2,\xi_2}^{1}(\theta_2)
  e^{i\lambda_2 \Delta\phi}\,.
\end{align}
where $\lambda_1$ and $\lambda_2$ denote the helicities of the two vector mesons.
The phase factor contains only the $\lambda_2$ value due to the choice of the reference frame.
Once the phase is factored out of the helicity coupling matrix $H_{\lambda_1,\lambda_2}$,
the symmetry relations for $H$ are significantly simpler as presented in the next section.

%

\subsection{Symmetry constraints} \label{sec:symmetries}
The matrix of the helicity couplings is strictly defined by
\begin{equation} \label{eq:helicity.def}
  H_{\lambda_1,\lambda_2} = \bra{JM;\lambda_1,\lambda_2}\hat{T}\ket{JM},
\end{equation}
where the bra-state is the projected two-particle state in the particle-2 phase convention,
the ket-state is the decaying state with the defined $J$ and $M$ in the GJ frame, and $T$ is the interaction operator~\cite{Martin:1970xx,Collins:1977jy}.
For brevity of notation we omit the index $J$ from the helicity matrix.
The following equations constrain the helicity couplings for specific $J^P$ sector.
The matrix is constrained by parity and permutation symmetry.
Parity transformation relates the opposite values of the vectors' helicities:
\begin{equation} \label{eq:parity}
  H_{\lambda_1,\lambda_2} = P (-1)^J H_{-\lambda_1,-\lambda_2},
\end{equation}
with $P$ being the internal parity of $X$.
The fact that the two vector mesons are identical relates the helicity matrix with the transposed one:
\begin{equation} \label{eq:permutation}
  H_{\lambda_1,\lambda_2} = (-1)^J H_{\lambda_2,\lambda_1}.
\end{equation}
The matrices of the helicity couplings are symmetric (anti-symmetric) for even (odd) spin $J$.
\begin{table*}[t]
  \caption{Possible quantum numbers of the decaying particle $X$ separated into four groups with respect to the symmetry of the helicity matrix. The framed quantum numbers in the last column have additional restrictions due to the maximal value of the spin projection.}
  \label{tab:couplings}
  \begin{ruledtabular}
    \begin{tabular}{c | r | r | l}
      group & signum $s=(-1)^{J}$ & naturality, $\epsilon = P(-1)^{J}$ & explicit $J^P$                    \\\hline
      \I    & even($+$)           & natural($+$)                       & \fbox{$0^+$}, $2^+$, $4^+$, $6^+$ \\
      \II   & even($+$)           & unnatural($-$)                     & \fbox{$0^-$}, $2^-$, $4^-$, $6^-$ \\
      \III  & odd($-$)            & natural($+$)                       & $1^-$, $3^-$, $5^-$, $7^-$        \\
      \IV   & odd($-$)            & unnatural($-$)                     & \fbox{$1^+$}, $3^+$, $5^+$, $7^+$
    \end{tabular}
  \end{ruledtabular}
\end{table*}
The relations in Eq.~\eqref{eq:parity} and Eq.~\eqref{eq:permutation} greatly reduce the number of free components of the helicity matrix, which can in general be written as
\begin{align} \label{eq:general.H}
  H & = \begin{pmatrix}
          b    & a           & c              \\
          s\,a & d           & \epsilon s\, a \\
          s\,c & \epsilon\,a & \epsilon\,b\end{pmatrix}
\end{align}
where $a$, $b$, $c$, and $d$ are the helicity couplings, $\epsilon = P(-1)^J$ is the naturality of $X$, and the signum, determined by whether the spin of $X$ is odd or even, is given by $s=(-1)^J$.
The value of $b$ is non-zero only for positive $s$, while non-zero $d$ requires both $s$ and $\epsilon$ to be positive.
According to the values of $\epsilon$ and $s$, all possible quantum numbers $J^P$ are split into four groups as shown in Table~\ref{tab:couplings}.
The helicity matrix for each groups is
\begin{align} \label{eq:matrices}
  \I\Rightarrow   & \begin{pmatrix}
                      b & a & c \\
                      a & d & a \\
                      c & a & b
                    \end{pmatrix}, &
  \II\Rightarrow  & \begin{pmatrix}
                      b & a  &    \\
                      a &    & -a \\
                        & -a & -b
                    \end{pmatrix}, & \\
  \III\Rightarrow & \begin{pmatrix}
                         & a &    \\
                      -a &   & -a \\
                         & a &
                    \end{pmatrix}, &
  \IV\Rightarrow  & \begin{pmatrix}
                         & a  & c \\
                      -a &    & a \\
                      -c & -a &
                    \end{pmatrix}.
\end{align}
In general, $a$, $b$, $c$, and $d$ are complex helicity couplings,
however several of them vanish for specific groups:
$c=d = 0$ for group $\II$; $b=c=d=0$ for group $\III$; and $b=d=0$ for group $\IV$.
There are three special cases for low $J^P$ where additional helicity couplings vanish due to the requirement $|\lambda_1-\lambda_2| \leq J$: $0^+$ in group $\I$, for which $a=c=0$;
$0^-$ in group $\II$ with $a=0$; and $1^+$ in group $\IV$ with $c=0$.

The helicity matrices of different groups are orthogonal to each other given the scalar product
\begin{align} \label{eq:sc.prod}
  (H_1\cdot H_2) = \mathrm{Tr}(H_1 H_2^\dagger).
\end{align}
They produce generally different angular distributions except for a few degenerate cases discussed below.
The scalar product in Eq.~\eqref{eq:sc.prod} is used to fix the normalization of $H$ and gives the relation between the helicity couplings:
\begin{align} \label{eq:norn}
  (H\cdot H) = 4|a|^{2}+2|b|^{2}+2|c|^{2}+|d|^{2} = 1.
\end{align}

The form of the helicity matrices in Eq.~\eqref{eq:matrices} immediately leads to the
conclusion of the Landau-Yang theorem~\cite{Yang:1950rg,Landau:1948kw}.
For the decay of $X$ to a pair of real photons,
$H_{0,\lambda} = H_{\lambda,0} = 0$, as the photon cannot carry the longitudinal polarization, $\lambda=0$.
Practically, this corresponds to setting to zero the second row and second column of the helicity matrix.
The matrix of group $\III$ completely vanishes,
hence, mesons with odd-natural $J^P$ cannot decay to two real photons. The special case of group $\IV$ with $c=0$ also vanishes so the decay of $J^P=1^+$ to two real photons is also forbidden.

\subsection{Angular modulations}

It is often convenient to simplify the problem and consider
only observables that are insensitive to the initial polarization.
Once the decay plane orientation is integrated over in Eq.~\eqref{eq:I.6},
the production polarization matrix $R$ collapses to its trace.
Indeed, using the properties of the Wigner D-function
and the normalization of $R$, $\mathrm{Tr}\,R = 1$ we find that
\begin{align} \label{eq:intensity.3}
  I(\theta_1,\theta_2,\Delta\phi) & = \int \frac{\diff \Omega\,\diff \phi_1}{8\pi^2} I(\Omega,\Omega_1,\Omega_2)                              \\\nonumber
                                  & = \sum_{\nu=-2}^{2} \sum_{\xi_1,\xi_2}^{\{-1,1\}}|A^{\nu}_{\xi_1,\xi_2}(\theta_1,\theta_2,\Delta\phi)|^2.
\end{align}
This intensity is a polynomial on trigonometric functions of the angles
with the coefficients determined by the helicity couplings.
For practical convenience we provide an explicit form of this expression
calculated for the general matrix $H$:
\begin{align}
  I(\theta_1,\theta_2, \Delta\phi) = \sum_{i=1}^{6} c_i f_i(\theta_1,\theta_2, \Delta\phi)
  \label{eq:practical}
\end{align}
where $f_i$ are normalized angular functions and $c_i$ are coefficients that depend on the helicity couplings.
The functional forms for $f_i$ and $c_i$ are given in Table~\ref{tab:harmonics}.
\begin{table*}[]
  \centering
  \caption{Basis functions and coefficients for the three-dimensional angular distribution $I(\theta_1,\theta_2,\Delta\phi)$ as expressed in Eq.~\eqref{eq:practical} for final states consisting of four leptons or four scalar particles.
  }
  \begin{ruledtabular}
    \begin{tabular}{r|r|r}
      $i$ & angular functions, $f_i$                                                                                                                                                & coeff. $c_i$ for $X\to V(l^+l^-)V(l^+l^-)$
      \\\hline
      $1$
          & $9\sin^{2}{\left (\theta_{1} \right )} \sin^{2}{\left (\theta_{2} \right )} \sin^{2}{\left (\Delta\phi \right )} /2$
          & $-\epsilon |b|^{2}/2$
      \\
      $2$
          & $\sin{\left (\theta_{1} \right )} \sin{\left (\theta_{2} \right )} \cos{\left (\theta_{1} \right )} \cos{\left (\theta_{2} \right )} \cos{\left (\Delta \phi \right )}$
          & $(9(\epsilon+1) \mathrm{Re}\,(b^* d) - 18 \epsilon |a|^{2} s)/4$
      \\
      $3$
          & $9\sin^{2}{\left (\theta_{1} \right )} \sin^{2}{\left (\theta_{2} \right )}/4$
          & $(2 \epsilon |b|^{2} - 8 |a|^{2} + 2 |b|^{2} + 2 |c|^{2} + 4 |d|^{2})/4$
      \\
      $4$
          & $3\sin^{2}{\left (\theta_{1} \right )}/2$
          & $(6 |a|^{2} - 3 (|b|^{2}+|c|^{2}))/2$
      \\
      $5$
          & $3\sin^{2}{\left (\theta_{2} \right )}/2$
          & $(6 |a|^{2} - 3 (|b|^{2}+|c|^{2}))/2$
      \\
      $6$
          & $1$
          & $9(|b|^{2}+|c|^{2})/2$
    \end{tabular}
  \end{ruledtabular}
  \label{tab:harmonics}
\end{table*}

There are potential cases where different hypotheses are not distinguishable.
If $a$ is the only non-zero helicity coupling,
the values of $\epsilon$ and $s$
enter the angular coefficients $c_2$ as the product $\epsilon s$,
and therefore
group-$\I$ has the same angular distributions as group-$\IV$,
while group-$\II$ is indistinguishable from group-$\III$.
Such vanishing of the helicity couplings, however,
is an exceptional case and might indicate some other symmetry or additional selection rule.

\begin{figure}
  \includegraphics[width=\linewidth]{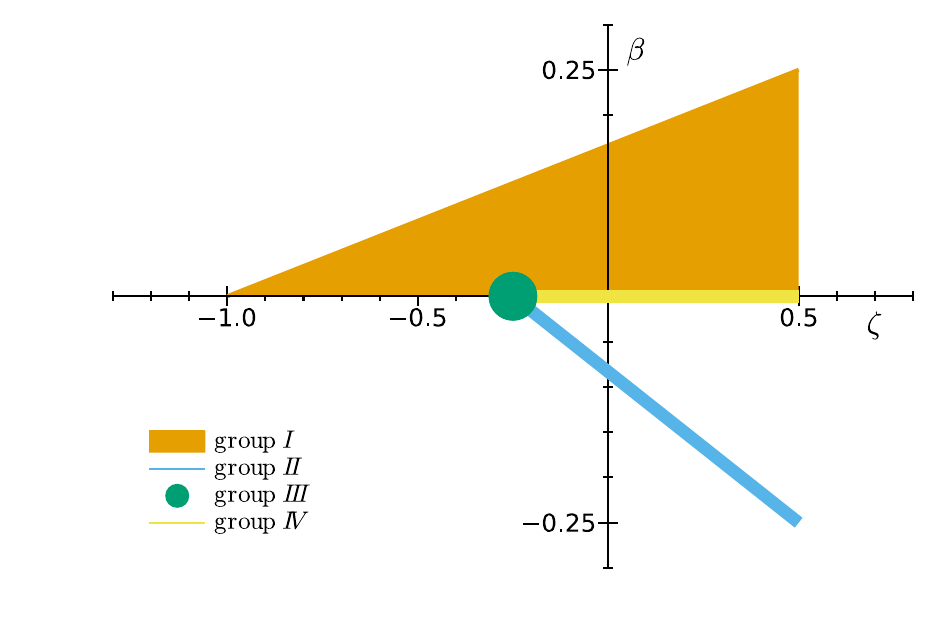}
  \caption{Allowed values of the angular modulations, $\beta$ and $\zeta$
    for the four spin-parity groups in the decays $X\to 4\mu$.}
  \label{fig:mu.beta.zeta.regions}
\end{figure}

One dimensional projections or moments can access the same quantities as can be seen from Eq.~\eqref{eq:practical} and Table~\ref{tab:harmonics}.
Integrating the intensity
over $\cos\theta_1$ and $\cos\theta_2$, the distribution of $\Delta\phi$ is given by
\begin{align} \label{eq:phi}
  I(\Delta\phi) = 1 + \beta \cos(2 \Delta\phi),
\end{align}
where $\beta = \frac{1}{2}\epsilon |b|^2/(4|a|^2+2|b|^2+2|c|^2+|d|^2)$.
For odd spins $\beta=0$, while for even spins,
the sign of the $\cos(2\Delta\phi)$ component in given by the parity.
The one-dimensional projection in $\cos\theta_i$, $i=1, 2$ also provides useful information on the helicity couplings:
\begin{align} \label{eq:theta}
  I(\cos\theta_i) & = 1 + \zeta\,\frac{3\cos^2\theta_i-1}{2}, \quad i=1,2.
\end{align}
with $\zeta = (|b|^2+|c|^2-|a|^2-|d|^2)/(4|a|^2+2|b|^2+2|c|^2+|d|^2)$.
The value of $\zeta$ always falls into the range $[-1,1/2]$.
Furthermore, $\zeta \geq -1/4$ for groups $\II$ and $\IV$ while it must be equal to $-1/4$ for group $\III$.
Fig.~\ref{fig:mu.beta.zeta.regions} summarizes the values possible for $\beta$ and $\zeta$ separated by group.

The idealistic situation of a single isolated resonance discussed so far is,
however, rarely achieved in practice:
resonances are usually produced superimposed on a continuum distribution or in the presence of additional states.
Moreover, interference effects modify the values of the angular asymmetries, $\beta$ and $\zeta$.
In such multicomponent production processes,
considerations of both the mass and decay angles of the system are essential to separate the different components.
The helicity matrix is modified to take into account the mass dependence of each component.
For a given spin $J$ and parity $P$, it is described by
\begin{align} \label{eq:decay.A2}
  H_{\lambda_1\lambda_2}^{J^P}(m) = \sum_{r}H_{\lambda_1\lambda_2}^{J^P,r} \mathcal{F}_r^{J^P}(m)\,,
\end{align}
where $H_{\lambda_1\lambda_2}^{J^P,r}$ is a matrix of the complex coupling parameters for the component $r$ having spin $J$ and parity $P$, $m$ is the mass, and $\mathcal{F}_r^{J^P}(m)$ is its lineshape function.
The resonance components have a pole in their parametrization leading to a fast variation
of the complex phase in $R_\pi$, while the
continuum production often follows a power law with slowly moving complex phase.

The amplitude factor that contributes to the intensity in Eq.~\ref{eq:I.6} must be summed over contributions with spins and parity,
and the intensity now has a mass dependence.
The angular asymmetries $\beta$ and $\zeta$ in the general case are given by the sum
of all components, where the terms with different quantum numbers are added
incoherently.
$\beta$ and $\zeta$ now depend on mass
with $\beta(m)=B(m)/D(m)$ and $\zeta=C(m)/D(m)$ where
\begin{widetext}
  \begin{align}\label{eq:betazetasum}
    B(m) & =
    \sum_{j\in \{J^P\}} \epsilon_j \left|\sum_{r} b^{r,j} F_{r}^j(m)\right|^2\,, \\ \nonumber
    C(m) & =
    2\sum_{j\in \{J^P\}}\left|\sum_{r} b^{j,r} \mathcal{F}_{r}^j(m)\right|^2
    +2\sum_{j\in \{J^P\}}\left|\sum_{r} c^{j,r} \mathcal{F}_{r}^j(m)\right|^2
    -2\sum_{j\in \{J^P\}}\left|\sum_{r} a^{j,r} \mathcal{F}_{r}^j(m)\right|^2
    -2\sum_{j\in \{J^P\}}\left|\sum_{r} d^{j,r} \mathcal{F}_{r}^j(m)\right|^2\,,
    \\  \nonumber
    D(m) & =
    4\sum_{j\in \{J^P\}}\left|\sum_{r} a^{j,r} \mathcal{F}_{r}^j(m)\right|^2
    +2\sum_{j\in \{J^P\}}\left|\sum_{r} b^{j,r} \mathcal{F}_{r}^j(m)\right|^2
    +2\sum_{j\in \{J^P\}}\left|\sum_{r} c^{j,r} \mathcal{F}_{r}^j(m)\right|^2
    +\sum_{j\in \{J^P\}}\left|\sum_{r} d^{j,r} \mathcal{F}_{r}^j(m)\right|^2\,,
  \end{align}
\end{widetext}
where the summation index $j$ counts all unique $J^P$ assignments and the index $r$ sums over those contributions with that given $J^P$.
The elements of the helicity matrix $H^{j,r}$ are denoted by $b^{j,r}$, $c^{j,r}$, $a^{j,r}$, $d^{j,r}$ according to Eq.~\eqref{eq:general.H}.

The expression for $D(m)$ describes the total mass spectrum, which, depending on the resonances present and their interference, may show peaking structures above or dips below the continuum.
The presence of narrow peaking structures are often taken as evidence for the presence of resonances.

The angular distributions $\beta(m)$ and $\zeta(m)$
present different projections of the intensity.
For the continuum, these will vary slowly as a function of mass.
They may even be constant if the relative $J^P$ components and components of the helicity matrix do not change with mass.
On the pole of a given resonance, it might be expected that $\beta$ and $\zeta$ give the values described above for isolated resonances, and if the resonance does not interfere with the continuum, then the observed value for $\beta$ or $\zeta$ on the pole will add to the value for the continuum and could be used to determine which spin-parity group the resonance belongs to.
However, in general this is not true and knowledge is required of the helicity structure of the continuum and its interference with the resonances.
However despite this limitation, it should be noted that if a resonance is present, the complex phase varies rapidly around the pole and this will often result in distributions for $\beta(m)$ and $\zeta(m)$ that also vary rapidly with mass.  This could be considered as evidence for the presence of resonances, in the same way that variations in the mass spectrum $D(m)$ do, and may help resolve ambiguities.

%


\section{Application} \label{sec:application}

\subsection{Spin analysis of an isolated resonance} \label{sec:higgs} 

The determination of spin and parity of an isolated resonance
is demonstrated using simulated data for the decay of the Higgs boson.
The form of the helicity matrix is found by considering the interaction term of the Higgs with the gauge bosons.
The covariant amplitude for \mbox{$H\to ZZ$} reads:
\begin{equation} \label{eq:HZZ}
  i\mathcal{M}^{H\to ZZ} = 2i\frac{m_Z^2}{v} (\varepsilon_1^*\cdot\varepsilon_2^*),
\end{equation}
where $m_Z$ is the mass of the $Z$ boson and $v$ is the vacuum expectation value of the Higgs field.
Using the explicit expressions for the polarization $\varepsilon_i(\lambda_i)$ vectors (see Appendix~\ref{sec:polarisation.vectors}),
the special form of the matrix for group $\I$ is
\begin{equation} \label{eq:H2ZZ}
  H^{H\to ZZ} = \frac{1}{\sqrt{3}}
  \begin{pmatrix}
    1 &    &   \\
      & -1 &   \\
      &    & 1
  \end{pmatrix}
  + O(p^2/m_Z^2),
\end{equation}
where $p = \sqrt{m_H^2 - 4m_Z^2}$ is the break-up momentum of the $Z$ boson in the rest frame of the Higgs boson.
The identity matrix corresponds to the $S$-wave in the decay, while the $D$-wave is proportional to $p^2$ and is suppressed at the $ZZ$ threshold. Furthermore, since one $Z$ must be virtual, there is a negligible contribution from the $D$ wave.
In the tests presented here, we consider the decay channel $H\to Z(e^+e^-)Z(\mu^+\mu^-)$;
for $Z$-boson decays to identical final states,
the interference between the two decay chains must be included.

\begin{figure}
  \includegraphics[width=0.46\textwidth]{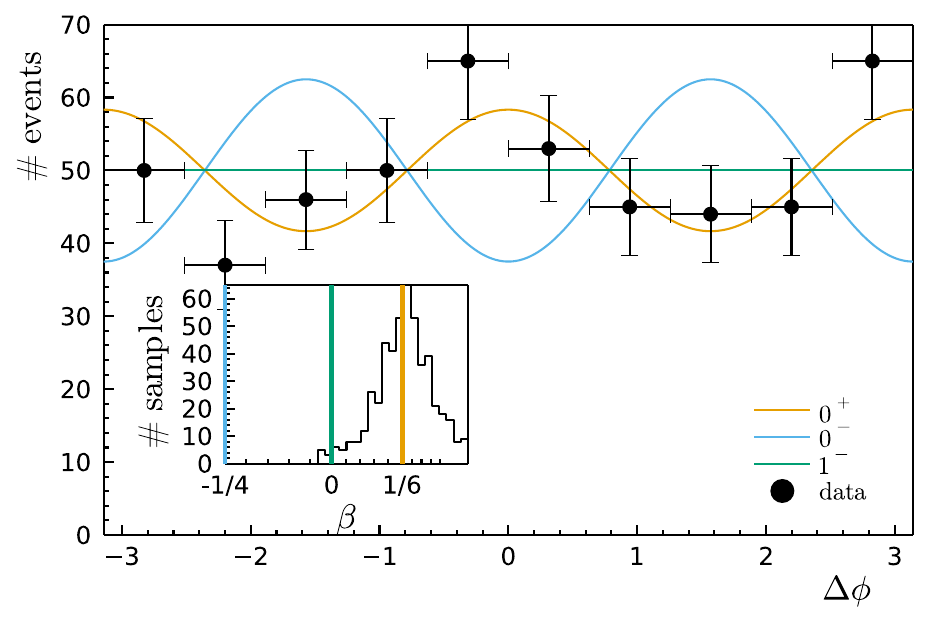}\\
  \includegraphics[width=0.46\textwidth]{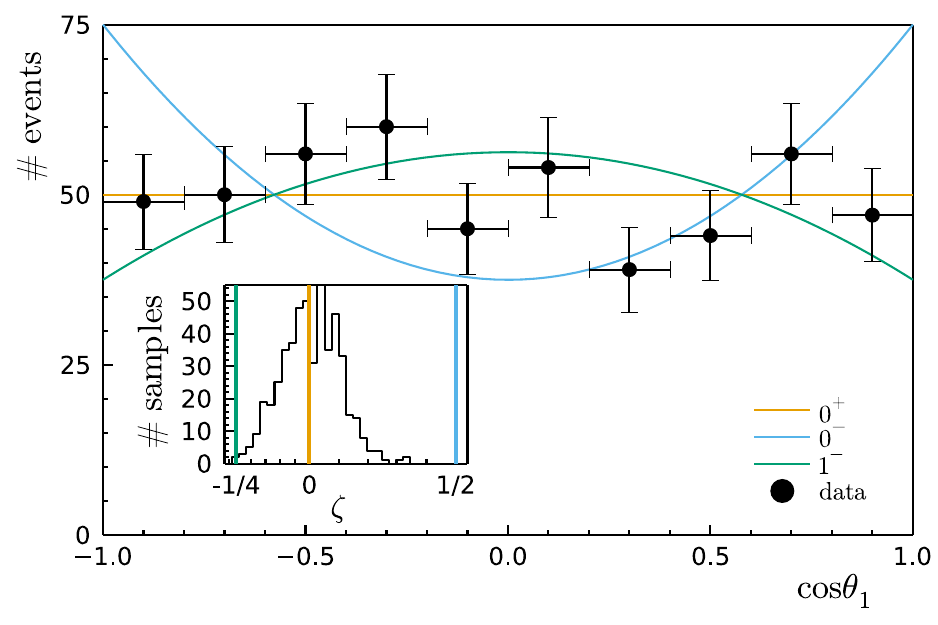}
  \caption{Distribution of the azimuthal angles $\Delta\phi$ (top) and $\cos\theta_1$ (bottom) for the Higgs decay to $e^+e^-\mu^+\mu^-$ with a single sample of $500$ simulated events generated with the $J^P=0^+$ hypothesis.
    The orange lines are the expectation curves under the $J^P=0^+$ hypotheses with $b=d=1/\sqrt{3}$.
    The blue lines give the expected dependence for the quantum numbers $J^P=0^-$ (group \II).
    The distributions for $J^P=1^-$ are shown by the green lines.
    The inset plots show distribution of the parameters $\beta$ and $\zeta$ for an ensemble of pseudoexperiments. The coloured lines indicate the true values corresponding to each hypothesis.
  }
  \label{fig:higgs.phi}
\end{figure}
A sample of $500$ simulated $J^P=0^+$ events, corresponding to the helicity matrix in Eq.~\eqref{eq:H2ZZ}, were generated using a dedicated framework written in \texttt{Julia}~\cite{julia.JpsiJpsi}.
Results from an analysis using the one-dimensional distributions given in Eq.~\eqref{eq:phi} and \eqref{eq:theta} are shown in
Fig.~\ref{fig:higgs.phi}.
Superimposed are the theoretical distributions expected
from $J^P=0^+,0^-$ and $1^-$; the data agree best with the $0+$ hypothesis
with the values $\beta_H = 1/6$ and $\zeta_H = 0$.
An ensemble of pseudoexperiments lead to estimations
for $\beta$ and $\zeta$ that are shown as inset plots
and compared to the
theoretical values for three spin hypothesises.
The width of the distributions indicate the precision
of the determination.
The $0^+$ and $1^-$ hypotheses are separated by about $2\sigma$,
while in the multidimensional approach the statistical significance is about $3.5\sigma$.
This shows the improvement in experimental precision that can be achieved by taking account of correlations between the angular variables.
Details on hypothesis testing are presented in Appendix~\ref{sec:test.statistics}.

\subsection{Structure of the $J/\psi J/\psi$ system}
\begin{figure*}
  \centering
  \includegraphics[width=0.9\textwidth]{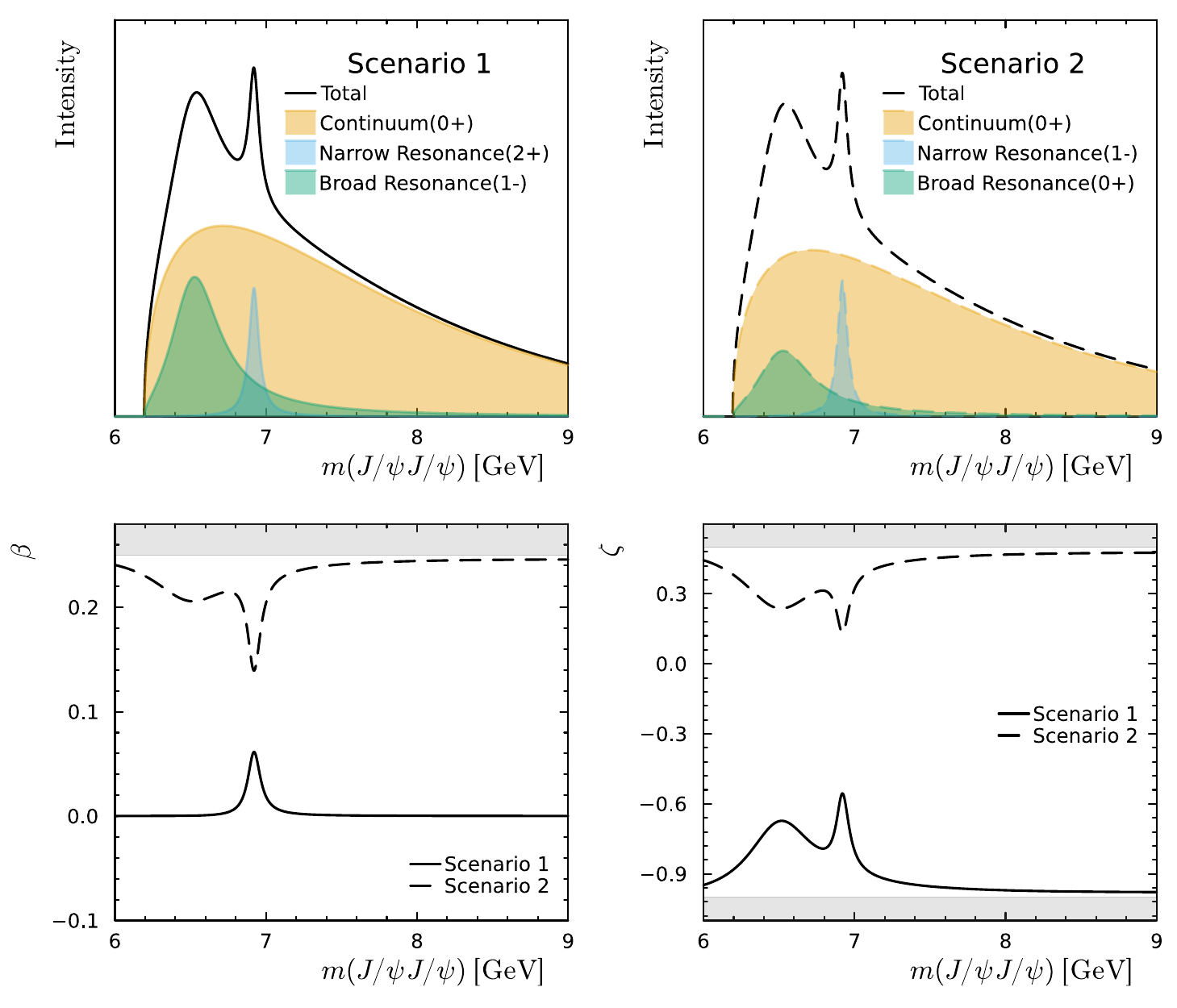}
  \caption{
    Top row: Intensity distributions for the two scenarios detailed in Table~\ref{tab:jjmodel}.
    Bottom row: The angular distributions for $\beta$ (left) and $\zeta$ (right) for each scenario.}
  \label{fig:jjtoy}
\end{figure*}

Examination of angular distributions can improve the understanding of the constituents of a mass spectrum, since
with multiple interfering components the situation is often unclear.
A typical example is the observation by LHCb~\cite{LHCb:2020bwg}, later confirmed by ATLAS~\cite{ATLAS:2023bft} and CMS~\cite{CMS:2023owd}, of structure in the $J/\psi J/\psi$ system that can be interpreted in several ways.
At LHCb, two models that have been used to fit the data consider:
firstly, one $T_{cc\bar{c}\bar{c}}(6900)$ resonance and two additional resonances above threshold that do not interfere with the continuum;
secondly, one $T_{cc\bar{c}\bar{c}}(6900)$ resonance that does not interfere with the continuum, and another broader resonance that might interfere.
We make a qualitative description of experimental spectrum in both models using an exponential function to describe the
continuum and Breit-Wigner functions for the resonances.
The parameters used are given in Table~\ref{tab:jjmodel} and the results are shown in Fig.~\ref{fig:jjtoy}.

\begin{table}[]
  \centering
  \caption{Model parameters used to simulate the $J/\psi J/\psi$ spectrum in two scenarious. For helicity couplings, only non-zero elements are listed. The resonance amplitude is parametrizatid by the Breit-Wigner formula, $m_0 \Gamma_0/(m_0^2 - m^2 - im_0\Gamma_0)$. The exponential shape, $e^{-\beta m}/u$ is used for the continuum, with the factor $u$ ensuring unit integral in the range below 3\,GeV above the threshold, $u = \beta (e^{3\beta}-1)e^{2m_{j/\psi}}$.
  } 
  \begin{tabular}{l|c|c}
                         & Scenario 1                       & Scenario 2          \\
    \hline
    \textbf{Continuum}   & \multicolumn{2}{c}{Exponential}                        \\
    Slope, $\beta$       & \multicolumn{2}{c}{0.425}                              \\
    $J^P$                & $0^+$                            & $0^+$               \\
    Helicity coupligns   & {$d=1$}                          & {$b=1/\sqrt{2}$}    \\
    Normalization        & 0.2                              & 0.2                 \\
    \hline
    \textbf{Resonance 1} & \multicolumn{2}{c}{Breit-Wigner}                       \\
    Mass, $m_0$          & \multicolumn{2}{c}{6920 MeV}                           \\
    Width, $\Gamma_0$    & \multicolumn{2}{c}{80 MeV}                             \\
    $J^P$                & {$2^+$}                          & {$1^-$}             \\
    Helicity coupligns   & {$b=d=1$}                        & {$a=1/2$}           \\
    Normalization        & {$2$}                            & {$2.2$}             \\
    \hline
    \textbf{Resonance 2} & \multicolumn{2}{c}{Breit-Wigner}                       \\
    $J^P$                & {$1^-$}                          & {$0^+$}             \\
    Helicity coupligns   & {$a=1/2$}                        & {$b=0.3i$, $d=0.9$} \\
    Mass, $m_0$          & 6500 MeV                         & 6492 MeV            \\
    Width, $\Gamma_0$    & 400 MeV                          & 450 MeV             \\
    Normalization        & {$2.5$}                          & {$1.84$}            \\
  \end{tabular}
  \label{tab:jjmodel}
\end{table}

The mass distributions are qualitatively similar and with this information alone it is difficult to distinguish between the two competing scenarios.
However, the angular distributions allow different structure to be resolved.
Far from the resonances the expected values of $\beta$ and $\zeta$ are determined for the continuum.
On the peaks of the resonances, deviations from these values are observed, although the nominal values expected for a pure resonance are not achieved due to an admixture of, and possible interference with, the continuum.
The structures in the angular distributions are correlated with those in the mass distributions and have widths related to the widths of the resonances.
In summary, the angular distributions provide different projections of the amplitude and in some cases may be sufficient to distinguish between different hypotheses, or assist with identification of the spin and parity of putative states.

\section{Conclusions}
\label{sec:conclude}
A framework has been presented for analysing resonances decaying into two identical vector mesons.
In the absence of information on the production polarization, the angular distributions for the decays are fully described through three decay angles.
Projections of these enable the decays to be characterised by two parameters, $\beta$ and $\zeta$ whose allowed values can be subdivided into four groups reflecting the $J^P$ of the parent.
Two examples of the use of this formalism were presented: one concerns the Higgs decays to $ZZ$ pairs; the other presents unknown resonances decaying to pairs of $J/\psi$ mesons.  The latter is an example of a frequent problem in spectroscopy when trying to determine the presence and quantum numbers of new resonances: consideration of the angular distributions can significantly aid in understanding the underlying physics.
We hope this work equips experimental groups with the tools required to incorporate angular distributions into their analyses and we encourage experiments to publish angular information for the decay products along with the mass distributions.

\section*{Acknowledgement}
The project was motivated by a discussion in the LHCb Amplitude Analysis group.
We thank Biplab Dey for organizing a meeting dedicated to $X\to VV$.
We would like to thank Alessandro Pilloni for useful comments on the work.

\pagebreak

\appendix
\onecolumngrid

\pagebreak

\section{Modifications for $X\to V(S^+S^-)V(S^+S^-)$} \label{sec:KKKK}
\begin{figure}[h]
  \includegraphics[width=0.5\linewidth]{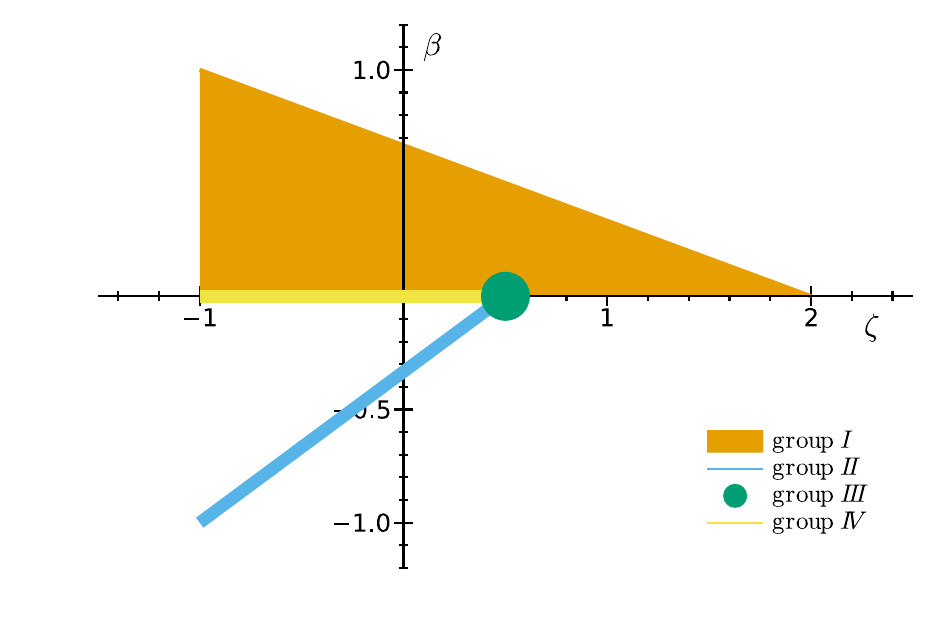}
  \caption{Allowed values of the angular modulations, $\beta$ and $\zeta$,
    for the four spin-parity groups in the decay of $X$ to four scalars.}
  \label{fig:K.beta.zeta.regions}
\end{figure}

The amplitude requires a small modification when a system of four scalar particles is considered,
e.g. $\rho(\pi^+\pi^-)\rho(\pi^+\pi^-)$ and $\phi(K^+K^-)\phi(K^+K^-)$.
The decay matrix element in Eq.~\eqref{eq:decay.A} reads:
\begin{align}
  A^{\nu}_{4S}(\theta_1,\theta_2,\Delta\phi) & = 3
  \sum_{\lambda_1,\lambda_2}
  \delta_{\nu,\lambda_1-\lambda_2}
  H_{\lambda_1\lambda_2}
  d_{\lambda_1,0}^{1}(\theta_1) d_{\lambda_2,0}^{1}(\theta_2)
  e^{i\lambda_2 \Delta\phi}
\end{align}
where the decay $V\to S^+S^-$ proceeds in $P$-wave only.
The functional forms for the three-dimensional angular distribution are given in Table~\ref{tab:harmonics_4k}.
The variation in the $\Delta\phi$-dependence and $\cos\theta_1$ is more pronounced
since there is no averaging over the spins of the final-state particles,
\begin{align}
  \beta_{(S)} & = 4\beta_{(l)} = \frac{2\epsilon|b|^2}{4|a|^2+2|b|^2+2|c|^2+|d|^2},                &
  \zeta_{(S)} & = -2\zeta_{(l)} = \frac{-2|b|^2-2|c|^2+2|a|^2+2|d|^2}{4|a|^2+2|b|^2+2|c|^2+|d|^2},
\end{align}
with lower indices of $(S)$ and $(l)$ to indicate scalar and lepton particles, respectively.
The correlation of the $\beta$ and $\zeta$ observables for different groups are shown in Fig.~\ref{fig:K.beta.zeta.regions}.

\begin{table}[h]
  \centering
  \caption{Basis functions and coefficients for the three-dimensional angular distribution $I(\theta_1,\theta_2,\Delta\phi)$ as expressed in Eq.~\eqref{eq:practical} for final states consisting of four leptons or four scalar particles.
  }
  \begin{ruledtabular}
    \begin{tabular}{r|r|r}
      $i$ & angular functions, $f_i$
          & coeff. $c_i$, for $X\to V(S^+S^-)V(S^+S^-)$                                                                                                                             \\\hline

      $1$
          & $9\sin^{2}{\left (\theta_{1} \right )} \sin^{2}{\left (\theta_{2} \right )} \sin^{2}{\left (\Delta\phi \right )} /2$
          & $- 2 \epsilon |b|^{2}$
      \\

      $2$
          & $\sin{\left (\theta_{1} \right )} \sin{\left (\theta_{2} \right )} \cos{\left (\theta_{1} \right )} \cos{\left (\theta_{2} \right )} \cos{\left (\Delta \phi \right )}$
          & $9(\epsilon+1) \mathrm{Re}\,(b^* d) - 18 \epsilon |a|^{2} s$
      \\

      $3$
          & $9\sin^{2}{\left (\theta_{1} \right )} \sin^{2}{\left (\theta_{2} \right )}/4$
          & $2 \epsilon |b|^{2} - 8 |a|^{2} + 2 |b|^{2} + 2 |c|^{2} + 4 |d|^{2}$
      \\

      $4$
          & $3\sin^{2}{\left (\theta_{1} \right )}/2$
          & $6 |a|^{2} - 6 |d|^{2}$
      \\

      $5$
          & $3\sin^{2}{\left (\theta_{2} \right )}/2$
          & $6 |a|^{2} - 6 |d|^{2}$
      \\

      $6$
          & $1$
          & $9 |d|^{2}$
    \end{tabular}
  \end{ruledtabular}
  \label{tab:harmonics_4k}
\end{table}

\section{Polarization vectors} \label{sec:polarisation.vectors}
To translate the covariant expression in Eq.~\eqref{eq:HZZ} to a helicity amplitude,
the explicit expressions for the polarization vectors are used:
\begin{align}
  \varepsilon_z^{\mu}(\pm1) & = \frac{1}{\sqrt{2}} \left( 0,\mp 1,-i,0 \right), &
  \varepsilon_z^{\mu}(0)    & = \frac{1}{m_Z} \left(p,0,0,E\right),
\end{align}
where $E$, $p$, and $m_Z$ are the energy, momentum, and mass of the $Z$ boson, respectively.
The general expressions for the rotational vectors follow:
\begin{align}
  \varepsilon_1(\lambda) & = R_z(\phi) R_y(\theta) \varepsilon_z(\lambda),                           \\
  \varepsilon_2(\lambda) & = (-1)^{1-\lambda} R_z(\phi) R_y(\theta) R_y(\pi) \varepsilon_z(\lambda),
\end{align}
where $R_y(\phi)R_y(\theta)$ is a product of the three-dimensional rotation matrices
that transforms the vector $(0,0,1)$ to the direction $(\sin\theta\cos\phi,\,\sin\theta\sin\phi,\,\cos\theta)$.
The second particle obtains an additional rotation by $\pi$ about the $y$ axis since we
use the particle-2 phase convention.

\section{Testing hypotheses} \label{sec:test.statistics}
Given a set of data corresponding to the decay of particle $X$, the questions arises as to how well its spin and parity can be determined using the angular distributions.
The most powerful method for testing a spin-parity hypotheses is a multidimensional fit, which takes into account the correlations between the angular variables.
We confine ourselves here to a three dimensional analysis removing the polarization degrees of freedom, although the discussion can be generalized to the full six-dimensional space treating the polarization as model parameters.
To determine which group the particle belongs to, we define a test statistic
\begin{align} \label{eq:test.statistics}
  \TS_{G/G'} = \LLH_G - \LLH_{G'},
\end{align}
where $\LLH_G$ is the maximized value of the averaged log likelihood
for group $G$ and is given by
\begin{equation} \label{eq:likelihood}
  \LLH_G = \frac{1}{N} \sum_{e=1}^{N} \log I((\theta_1,\theta_2,\Delta\phi)_e|G\{\hat{h}\})
\end{equation}
where the sum runs over the $N$ events in the sample.
The intensity is calculated for each event,
assuming it belongs to group $G$ with the helicity couplings $\hat{h}$ that maximize the likelihood.

\begin{figure}
  \includegraphics[width=0.48\textwidth]{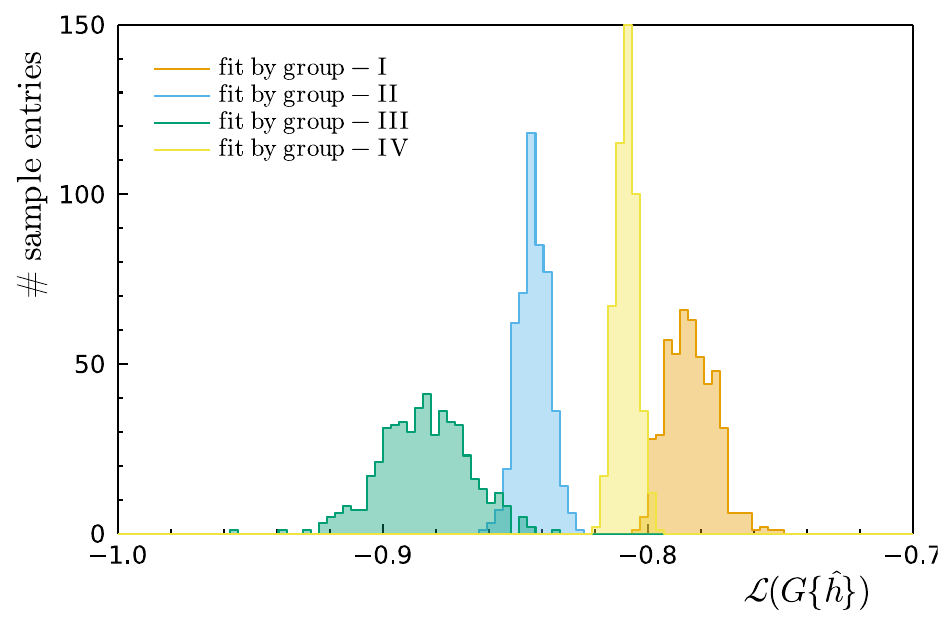}
  \includegraphics[width=0.48\textwidth]{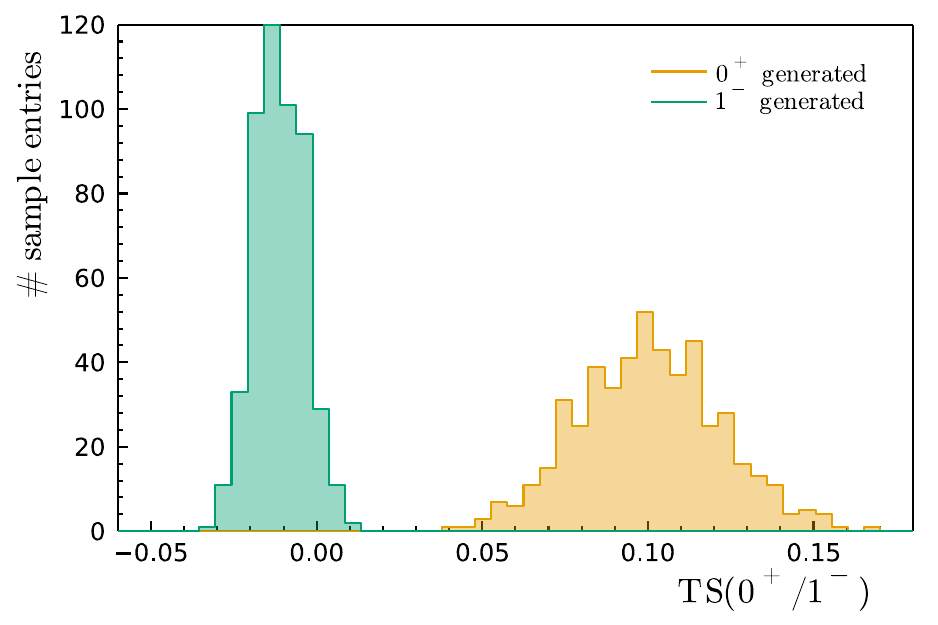}
  \caption{Left: Distribution of the
    maximized log-likelihood functions in a series of pseudoexperiments,
    each of which consists of 500 generated $H\to Z(e^+e^-)Z(\mu^+\mu^-)$ decays.
    Right: Distribution of the test statistic $\TS(0^+|1^-)$ for two hypotheses, $J^P=0^+$ and $J^P=1^-$, in two sets of pseudoexperiments.
  }
  \label{fig:TS.H.LLH}
\end{figure}

Distributions of the test statistic in pseidoexperiments is shown in Fig.~\ref{fig:TS.H.LLH}, together with a distribution of the likelihood values for different hypotheses.

\bibliography{ref}
\end{document}